# Co-Channel Interference Cancellation in OFDM Networks using Coordinated Symbol Repetition and Soft Decision MLE CCI Canceler

Manar Mohaisen *Student Member, IEEE* and KyungHi Chang, *Senior Member, IEEE*
The Graduate School of IT and T, Inha University
253 Yonghyun-Dong, Nam-Gu, 402-751 Incheon, KOREA
Email: lemanar@hotmail.com, khchang@inha.ac.kr

*Abstract*—In this paper, a new scheme of downlink co-channel interference (CCI) cancellation in OFDM cellular networks is introduced for users at the cell-edge. Coordinated symbol transmission between base stations (BS) is operated where the same symbol is transmitted from different BS on different sub-carriers. At the mobile station (MS) receiver, we introduce a soft decision maximum likelihood CCI canceler and a modified maximum ratio combining (M-MRC) to obtain an estimate of the transmitted symbols. Weights used in the combining method are derived from the channels coefficients between the cooperated BS and the MS. Simulations show that the proposed scheme works well under frequency-selective channels and frequency non-selective channels. A gain of 9 dB and 6 dB in SIR is obtained under multipath fading and flat-fading channels, respectively.

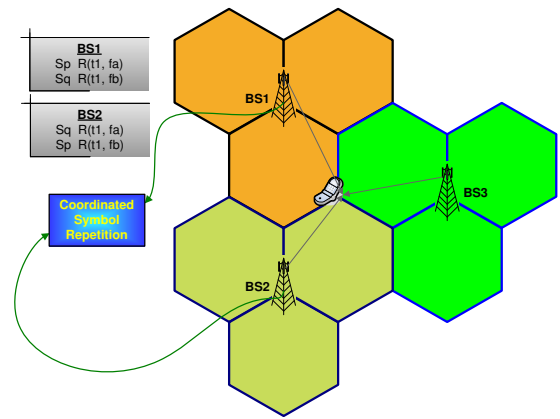

Fig. 1. OFDM cellular network with CCI and CSR.

## I. INTRODUCTION

Orthogonal frequency division multiplexing (OFDM) has gained attention for radio transmission technology of next generation mobile systems, due to its advantages including its robustness in multipath fading environment. However, the performance of OFDM cellular network is affected by the interference more than by any other performance limiting factor [1]. Furthermore, users at the cell edges in a fully-loaded system experience a high co-channel interference (CCI) that degrades the performance. Techniques have been proposed to mitigate the CCI which can be categorized into two categories, namely: interference avoidance and interference averaging. When coordinated interference avoidance are available, interference avoidance techniques such as adaptive interference coordination based on cell load can be applied. Otherwise, interference averaging techniques such as low rate forward error correction (FEC) and interleave division multiple access (IDMA) can be applied [2]. These techniques of CCI mitigation are effective when the system is not fully loaded. However, the performance of these techniques is highly degraded when the system is fully loaded [3]. On the other hand, multiple antenna techniques have been investigated to reduce the interference by exploiting the spatial signatures of the desired and the CCI signals [4].

In this paper, a sub-carrier-based virtual MIMO (V-MIMO) coordinated symbol repetition (CSR) is proposed for OFDM-based cellular systems to reduce CCI at the mobile station receiver. Two different sub-carriers are used to transmit the same symbol from two cooperated base stations (BS). Thus, the proposed scheme can be considered as a 2×2 MIMO system. In addition, we introduce a soft decision maximum-likelihood CCI canceler to extract the desired signal from the received signal. The outputs of the MLE CCI canceler are normalized using normalization factors calculated from the estimated channels between the mobile station (MS) and the cooperated BS to get channel diversity. The proposed CCI cancellation scheme can reduce effectively the CCI in frequency selective and non-selective fading channels without neither scarifying the full frequency reuse nor increasing the complexity of the receiver of the MS.

This paper is organized as follows. In section II, we introduce the effect of CCI on the BER performance degradation in OFDM cellular network. Section III presents the proposed CCI cancellation scheme with the modified MS receiver structure and the soft-decision CCI canceler. Section IV presents the simulation results and finally we draw conclusions in section V.

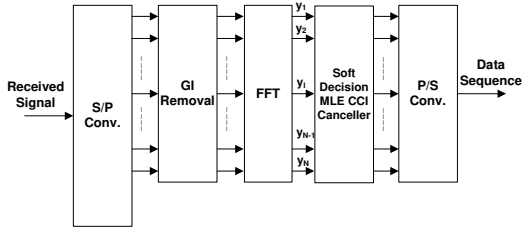

Fig. 2. MS receiver structure.

## II. Effect of CCI on the BER Performance Degradation in OFDM Cellular Networks

Herein, we present a general formula of the effect of CCI on the BER performance degradation in OFDM cellular networks. For detailed derivation refer to [5].

After a long derivation, the BER performance in OFDM cellular network with CCI for QAM modulation is given by

$$P_e = \frac{1}{log_2(\sqrt{M})} \cdot \left(1 - \frac{1}{\sqrt{M}}\right) \cdot \left[1 - \frac{1}{\sqrt{\frac{M-1}{3}\left[\sum_{j=2}^{K+1}\frac{1}{SIR_j} + \frac{2}{(log_2(M))\frac{E_b}{N_0}}\right] + 1}}\right] \quad (1)$$

where $M$ is the QAM modulation order, $K$ is the number of CCI interferer BS, $SIR$ is the signal-to-interferer ratio and $E_b/N_0$ is the average signal-to-noise (SNR) ratio per bit.

## III. System Model

In cellular OFDM system, the received signal at the MS is the sum of the desired signal and the CCI signals from other BS, as shown in Fig. 1. The received baseband signal at the MS receiver is given by

$$y(l) = \sum_{i=1}^{N_{BS}} h_i(l) s_i(l) + n(l) \quad (2)$$

where $s_i(l)$ and $h_i(l)$ are the transmitted signal and the channel coefficient between the $i^{th}$ BS and MS, at the $l^{th}$ sub-carrier, respectively. Furthermore, $n(l)$ is the double sided AWGN with zero mean and variance $\sigma_n^2$ and $N_{BS}$ is the number of BS. In a three-cell scenario, the data of the MS at the cell-edge is transmitted from two different BS, the transmission matrix can be written as follows

$$\begin{array}{c} \\ f_a \\ f_b \end{array} \begin{array}{cc} BS1 & BS2 \\ \left[ \begin{array}{cc} s_p & s_q \\ s_q & s_p \end{array} \right] \end{array} \quad (3)$$

where $f_a$ and $f_b$ are the $a^{th}$ and $b^{th}$ sub-carriers, respectively. Also, $s_p$ and $s_q$ are the transmitted symbols while $BS1$ and $BS2$ are the cooperated BSs. In addition, $BS3$ transmits a CCI signal. Here, the overall $N$ data sub-carriers are divided into $G$ groups each composed of $M$ sub-carriers.

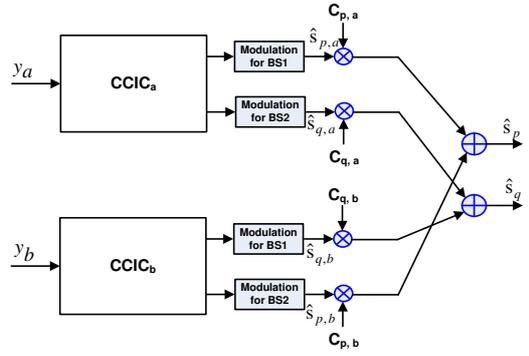

Fig. 3. Soft decision MLE CCI canceler.

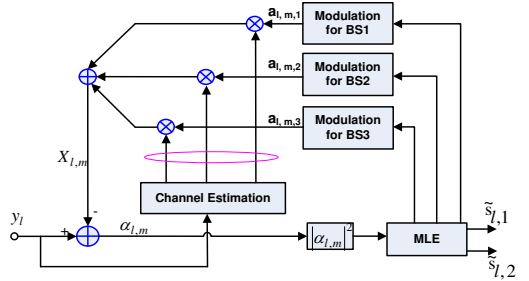

Fig. 4. CCIC.

The cooperation between BS is applied over groups with fluctuating or bad reported $SINR$ to improve the system performance without scarifying the spectral efficiency.

Fig. 2 shows the MS receiver structure; the received signal is firstly serial to parallel converted, the cyclic prefix (CP) is removed and then FFT is applied to get the frequency domain signal. At each sub-carrier, the soft decision MLE CCI canceler is applied to reduce the interference.

Fig. 3 depicts the structure of the soft decision MLE CCI canceler. At first, the hard decisions are obtained by passing the received signals on each sub-carrier into the MLE CCI canceler (CCIC). Only the estimated signals of the cooperated BS are passed into the next step where the interference signals are not considered. The CCIC generates replicas of the transmitted signals by computing all the weighted possible combinations as shown in Fig. 4. Herein, the weights represent the estimated channel coefficients at every data sub-carrier. The replica with the minimum Euclidean distance is chosen and the hard decisions are obtained [6]. The square Eucledian distance between received signal and generated replicas can be written as

$$|\alpha_{l,m}|^2 = \left| y(l) - \sum_{i=1}^{3} \hat{h}_i(l) a_i(l,m) \right|^2 \quad (4)$$

where $y(l)$ and $\hat{h}_i(l)$ are the received signal and the estimated channel coefficient between the $i^{th}$ BS and the MS at the $l^{th}$ sub-carrier, respectively. Also, $a_i(l,m)$ are locally generated symbols to construct the replica $X_{l,m}$ where $m$ is the replica index. Finally, the estimated symbols at the sub-carriers $f_a$ and $f_b$ (i.e. $\hat{s}_{p,a}$, $\hat{s}_{p,b}$, $\hat{s}_{q,a}$ and $\hat{s}_{q,b}$) are weighted according to the channel coefficients over which they are transmitted then combined. The soft decisions at the output of the soft decision MLE CCI canceler are given by

$$\hat{s}_p = \frac{\left|\hat{h}_1(a)\right|}{2\left(\left|\hat{h}_1(a)\right|+\left|\hat{h}_2(a)\right|\right)}\hat{s}_{p,a} + \frac{\left|\hat{h}_2(b)\right|}{2\left(\left|\hat{h}_1(b)\right|+\left|\hat{h}_2(b)\right|\right)}\hat{s}_{p,b} \quad (5)$$

and

$$\hat{s}_q = \frac{\left|\hat{h}_2(a)\right|}{2\left(\left|\hat{h}_1(a)\right|+\left|\hat{h}_2(a)\right|\right)}\hat{s}_{q,a} + \frac{\left|\hat{h}_1(b)\right|}{2\left(\left|\hat{h}_1(b)\right|+\left|\hat{h}_2(b)\right|\right)}\hat{s}_{q,b} \quad (6)$$

where $\hat{h}_i(l)$ is the channel coefficient between the $i^{th}$ BS and the MS at the $l^{th}$ sub-carrier. This combining given by equations 5 and 6 can be considered as a modified maximum ratio combining (M-MRC) method [7] where received symbol with high respective power has a high weight in the combining scheme while received symbol with low respective power has a low weight.

## IV. SIMULATION RESULTS

Performance of the proposed soft decision MLE CCI canceler with CSR (i.e. V-MIMO) is evaluated for users at the cell edge. To detect the transmitted symbol at every sub-carrier, $M^K$ replicas are calculated where $M$ is the modulation order and $K$ is the number of BS. For example, for QPSK modulation ($M = 4$) and two CCI signals, the MS receiver should calculate $4^3 = 64$ replicas to detected the transmitted symbol. This indicates that the overall complexity of the proposed algorithm is acceptable compared to other CCI cancellation schemes. Three BS are considered where CSR is applied between BS1 and BS2. The total number of sub-carriers, the length of the CP, and modulation order used in the simulations are 64, 16 and 4 (QPSK), respectively. In addition, no forward error correction (FEC) is applied. Furthermore, when the BER performance is investigated for different average $E_b/N_0$ values, average signal to interference ratio between BS1 and BS2 ($SIR_{12}$) and that between BS1 and BS3 ($SIR_{13}$) are set to 0 dB and 10 dB, respectively. Otherwise, average $E_b/N_0$ and $SIR_{12}$ are fixed to 18 dB and 0 dB, respectively. Also, a frame is composed of 57 OFDM symbols; 51 data symbols and 6 pilot symbols used for channel estimation where pilot patterns of the three BS are mutually orthogonal. Table I shows the main simulation parameters.

TABLE I
SIMULATION PARAMETERS.

| Parameter | Value |
|---|---|
| Carrier Frequency | 2 GHz |
| Number of Cells | 3 |
| Bandwidth | 20 MHz |
| FFT Size | 64 |
| Guard Interval | 16 |
| Modulation | QPSK |
| Speed | 10 Km/h |

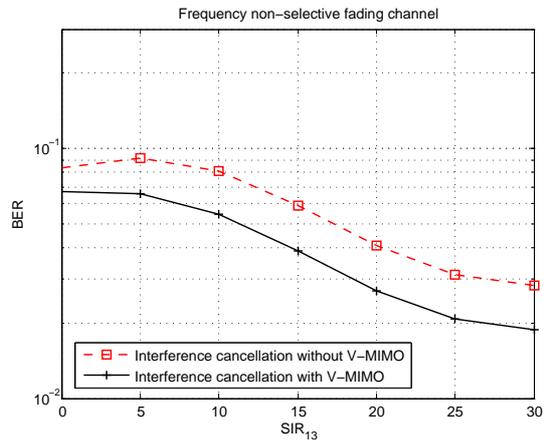

Fig. 5. SIR improvement of proposed scheme under frequency non-selective channel.

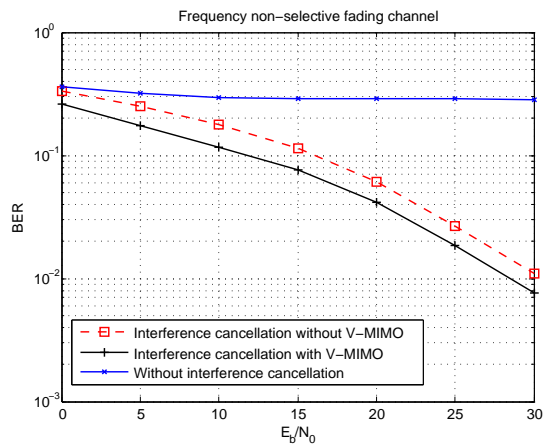

Fig. 6. $E_b/N_0$ improvement of proposed scheme under frequency non-selective channel.

### A. Simulation Under Frequency Non-Selective Fading Environment

In this subsection, single-path Rayleigh fading channel is considered. Unlike the conventional symbol repetition method which does not work well in low frequency selectivity channels, the proposed algorithm works well under frequency non-selective channels. When the same symbol is transmitted from two different BS, the instantaneous channel power between BS and MS are not constant. Thus, one can get channel power diversity. Fig. 5 shows the improvement of the proposed scheme; at BER of $5 \times 10^{-2}$, we get a gain of 6 dB in SIR. Fig. 6 shows the BER performance versus $E_b/N_0$ of the proposed scheme. In addition, the BER performance without any kind of CCI cancellation nor coordinated transmission between BS, presented in section II, is also shown. Furthermore, the dashed line shows the BER without coordination between BS. At BER of $5 \times 10^{-2}$, we have a gain of 3 dB compared to the system

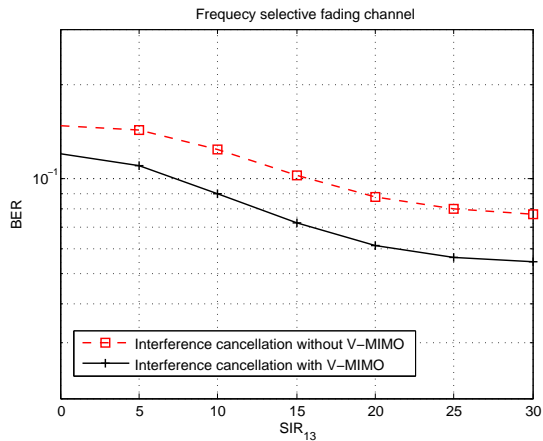

Fig. 7. SIR improvement of proposed scheme under frequency selective channel.

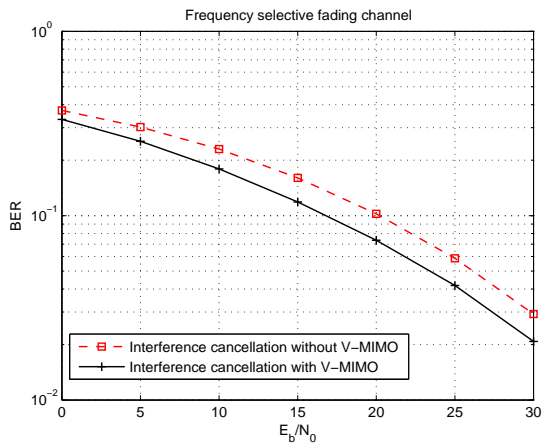

Fig. 8. $E_b/N_0$ improvement of proposed scheme under frequency selective channel.

performance without the proposed CCI cancellation scheme.

*B. Simulation Under Frequency Selective Fading Environment*

For multi-path fading environment, we consider a 5-path Rayleigh distributed fading channel. The power delay profile is a general exponential decay model with path interval of three samples durations. Fig. 7 shows the BER performance versus $SIR_{13}$, at BER of $9 \times 10^{-2}$, we obtain a gain of 9 dB in the SIR. Fig. 8 shows the BER performance of the proposed scheme at fixed average SIR values. We obtain a gain of 3.3 dB in the average $E_b/N_0$ at BER of $6 \times 10^{-2}$. A slight additional improvement is obtained under multipath fading channel because of the frequency diversity that is achieved by transmitting the same symbol on different sub-carriers from different BS.

## V. CONCLUSION

In this paper, we introduced the concept of using V-MIMO (by applying CSR) in CCI cancellation in fully-loaded OFDM networks. In addition, we introduced the construction of soft decision MLE CCI canceler and a modified MRC scheme to combine the softly decided symbols transmitted from different BS on different sub-carriers. The proposed scheme works well in frequency-selective environment by obtaining frequency diversity. Also, the proposed scheme works well in frequency non-selective fading channels by obtaining the short duration power diversity of the channels between the cooperated BS and the MS. Simulation results showed that the proposed scheme can efficiently reduce degradation caused by CCI and as consequence the system performance is improved.